\documentstyle[prl,amssymb,aps,epsfig,epsf,multicol]{revtex}

\begin{document}

\draft

\title{Nonuniversal behavior of scattering between fractional quantum Hall
edges}

\author{Bernd Rosenow$^{1,2}$ and Bertrand I. Halperin$^2$}

\address{ $^1$ Institut f\"ur Theoretische Physik, Universit\"at zu K\"oln,
D-50937 K\"oln, Germany   \\
$^2$ Physics Department, Harvard University, Cambridge,
MA 02138, USA}

\date{\today}

\maketitle

\begin{abstract}
  Among the predicted properties of fractional quantum Hall states are
  fractionally charged quasiparticles and conducting edge-states
  described as chiral Luttinger liquids.  In a system with a narrow
  constriction, tunneling of quasi-particles between states at
  different edges can lead to resistance and to shot noise. The ratio
  of the shot noise to the backscattered current, in the weak
  scattering regime, measures the fractional charge of the
  quasi-particle, which has been confirmed in several experiments.
  However, the non-linearity of the resistance predicted by the chiral
  Luttinger liquid theory was apparently not observed in some of these
  cases. As a possible explanation for these discrepancies, we
  consider a model where a smooth edge profile leads to formation of
  additional edge states. Coupling between the current carrying edge
  mode and the additional phonon like mode can lead to {\it
    nonuniversal} exponents in the current-voltage characteristic,
  while preserving the ratio between shot noise and the back-scattered
  current, for weak backscattering. For special values of the
  coupling, one may obtain a linear I-V behavior.
\end{abstract}

\pacs{PACS numbers: 71.10.Pn, 73.43.-f, 73.43.Jn}


\begin{multicols}{2}
  
The physics of interacting charge carriers confined to two spatial
dimensions and subject to a magnetic field has turned out to be
surprisingly rich and continues to pose ever new questions and to
unveil new phenomena. In the regime of the fractional quantum Hall effect
(FQHE), the effect of interactions is so strong that the bulk of the
system turns into an incompressible quantum liquid and the edges
carry strongly interacting excitations described as a chiral
Luttinger liquid (LL) \cite{review}. In addition, the elementary
excitations themselves are very different from the original charge
carriers in that they have a fractional charge \cite{Laughlin83}.

The fractional charge of FQHE quasi-particles can be detected in shot
noise measurements \cite{KaFi94} .  While the resistance of a single
chiral edge state is not renormalized by scattering centers, as the
excitations in a chiral LL move in one direction only, scattering from
one edge of a sample to the other edge induces both resistance and
shot noise.  In the weak scattering limit, scattering events are
independent of each other and are described by a Poisson process.
Then, the variance of the number of scattering events is equal to
their average, and the strength of current fluctuations $\langle
\left( \Delta I \right)^2$ is proportional to the product of
backscattered current and  the quasi-particle charge.

shot noise experiments carried out by several groups have confirmed
the predicted fractional charge of FQHE particles at filling fractions
$\nu=1/3$ and $\nu=2/5$
\cite{PiReHeUm97,RePiGrHe99,SaGlJi97,GrCoHe00}.  In at least some of
these cases, however \cite{PiReHeUm97,RePiGrHe99}, it appears that the
samples did not exhibit the strongly nonlinear resistance expected for
scattering between chiral LLs \cite{KaFi92}.  In this letter we
suggest that the scattering of fractionally charged quasi-particles may
indeed give rise to a non-universal current-voltage characteristic
over an interesting range, while preserving the ratio of shot noise to
backscattered current, if the chiral edge states couple to additional
degrees of freedom.  In the presence of a smooth confining potential,
edge reconstruction may occur \cite{Mac90,HaLe96,LeShHa00} and give
rise to additional phonon like modes coupling to the original chiral
LL.  In our model we assume that there is coulomb coupling between the
additional modes and the chiral edge states, but there is no tunneling
of charge between the modes, or between the additional modes and the
contacts.  The additional modes therefore carry no net electric
current, but can change the dynamics of backscattering between the two
edges.

\begin{figure}
\vspace*{.5cm}
\epsfig{file=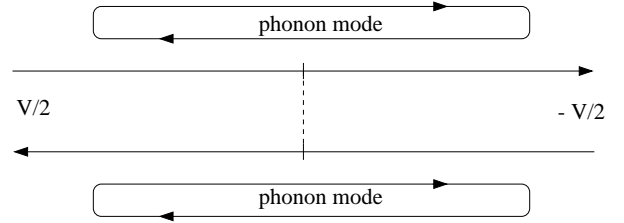,width=8cm}
\vspace*{.5cm}
\caption{Smooth edge profile: each chiral mode is coupled to a phonon-like 
mode with excitations moving in both directions. The dotted line is a 
point of tunneling between the two chiral edges.}
\label{coupling.f}
\end{figure}

Specifically, we discuss the following setup. From two reservoirs at
voltage $\pm V/2$ current is injected into the chiral edges at the
boundary of an incompressible quantum Hall state with filling factor
$\nu$.  We specialize on a simple situation with only one edge state
but expect that our general conclusions hold true in more complex
situations as well.  In the center of the sample, a constriction in
the Hall bar allows for scattering of charge $\nu$ quasi-particles from
one edge to the other.

We assume that in a region of length $L$ about the constriction, there
exists at each edge a pair of extra branches of excitations, moving
both to the left and to the right, which can happen if the confining
potential is relatively soft.  For example, if there were a
one-dimensional (fluctuating ) Wigner crystal of electrons in the
low-density region between the edge of the bulk quantized Hall state
and the fully-depleted edge of the sample \cite{HaLe96}, then the
extra modes would arise as acoustic phonon modes of the electron
crystal.  Equivalently, one could envisage at each edge a very narrow
strip of electrons in a fractional Hall state with lower filling
fraction than the bulk \cite{Mac90,LeShHa00}. The extra modes in this
case would correspond to the left and right-moving edges of the
low-density strip.  In any case, we shall refer to the extra modes as
"phonon modes" throughout this paper.  The model geometry is indicated
schematically in Figure 1.

We assume that disorder can be ignored, as a first approximation, and we
ignore all scattering of electrons between the various modes, other than
backscattering at the constriction. The phonon mode at each edge is
coupled via a short-range density-density interaction to the charge on
the adjacent chiral edge state of the bulk system. (We assume that the
long-range part of the Coulomb interaction is screened by an adjacent metal
gate.) The phonon modes themselves may be characterized by the sound
velocity they would have in the absence of coupling to the chiral edge
state and by an effective Luttinger liquid interaction parameter determined
by the mass density and  one-dimensional compressibility; alternatively, the
Luttinger parameter can be characterized by an effective filling factor
$\bar{\nu}$ for the low-density strip, which we assume to be smaller than
the bulk $\nu$.

Our analysis shows that for
frequencies larger than $\omega_L=v_F/L$ ($v_F$ is the Fermi velocity
in the chiral states), this coupling renormalizes the LL parameter
$\nu$ of the chiral edge to $\nu F > \nu$. The strength of this
renormalization depends on the coupling and on the ratio of the
velocities of chiral and phonon modes. As a result, the scattering
dynamics for voltages and temperatures larger than $\omega_L$ is
described by a {\it nonuniversal exponent} depending on the
interaction strength.

For frequencies smaller than $\omega_L$, this coupling is not
effective and the dynamics stays unchanged. In the experiments, shot
noise was measured at frequencies in the kilohertz or low megahertz
range; we assume that this is small compared to $\omega_L$.  As a
result, the quasi-particle charge is not renormalized and can be
extracted from a noise measurement.  In the limit $V \gg T$, it is
given by the ratio of shot noise and backscattered current.

We describe the chiral edges and scattering between them by the
finite temperature action

\begin{eqnarray}
{\cal S} &=&{\hbar \over 2 \nu} \int\!\! dx
dx^\prime d\tau d\tau^\prime \Theta(x,\tau) \, G^{-1}(x,x^\prime;\tau-
\tau^\prime) \, \Theta(x^\prime,\tau^\prime)\nonumber\\
& & \hspace*{-.5cm} + \lambda \int\!\! d\tau 
\cos\left[2 \sqrt{\pi} \Theta(0,\tau)
\right]\ . 
\label{totalaction}
\end{eqnarray}

Here, $\nu$ is the filling factor of the bulk with $1/\nu$ an odd
integer, and the inverse Green function $G^{-1}$ describes the
dynamics of the chiral edge states in the presence of the phonon
modes.  As the coupled system is quadratic in all densities, $G$ can
be calculated exactly. The left and right moving charge densities are
given by $\rho_\pm(x,\tau) =1/\sqrt{\pi} (\partial_x \pm (i/v_F)
\partial_\tau)\Theta(x,\tau)$, and the coupling to the reservoirs is
described by radiative boundary conditions \cite{EgGr98}

\begin{eqnarray}
\rho_\pm(\pm \infty,\tau)= \pm \nu e V / 2 \pi \hbar v_F\ .
\end{eqnarray}

As we are interested in calculating quasi-particle scattering, we
integrate over all fields away from the impurity site and obtain an
effective action for $\Theta(x=0,\tau)$.  Its dynamics is governed by
$G(0,0;\omega_n)$, where $G$ is the Green function of the
$\Theta$-field coupled to a phonon system of length $L$. We have
calculated $G$ in real space as a solution of the saddle point
equation with a delta-function inhomogeneity. We find that it shows a
crossover between two continuum theories: for low-frequencies the coupling to
the phonons is not effective, whereas in the high-frequency regime one finds
the same dynamics as for an infinite length phonon system, 

\begin{eqnarray}
\label{Greencross}
& &\hspace*{.5cm} { \nu \over 2  |\omega_n|}
\ \ \ \ \ \ \   , \, |\omega_n| \ll v_F /L\\  
G(0,0;\omega_n)&\sim&  \nonumber   \\
& & \hspace*{.5cm} { \nu F(u,\gamma) \over 2  |\omega_n|} \ \  ,\,  |\omega_n| \gg v_F /L \ . 
\nonumber
\end{eqnarray}

\vspace*{.5cm}
\begin{figure}
\centerline{
\epsfig{file=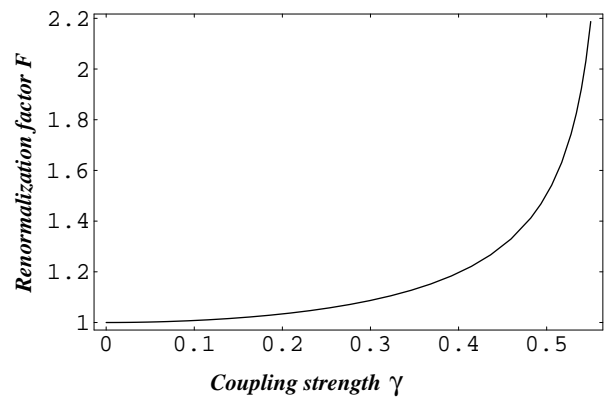,width=8cm}
}
\hspace*{.5cm}
\caption{Dynamical renormalization factor $F(u,\gamma)$ plotted against
the coupling strength $\gamma$ for the special value of the phonon velocity
$u=0.61$.}
\label{function.f}
\end{figure}

The renormalization factor $F(u,\gamma)$ is most easily calculated by 
integrating out the phonon systems in a continuum theory, 

\begin{eqnarray}
  \label{eq:F}
F(u,\gamma)&=& \int_{-\infty}^\infty
  {dx\over \pi} {1- \gamma^2 {2 u x^2\over 1 + u^2
        x^2}
\over {1  + x^2 \left(1 - \gamma^2 {2 u x^2 \over 1 + u^2
        x^2}\right)^2}}\ ,  
\end{eqnarray}
with $\gamma^2={\nu \tilde{\nu} \kappa^2 \over 2 \pi^2}$.  Here,
$\kappa$ is the coupling constant of the density-density interaction
between a phonon mode and the adjacent chiral mode, and $u$ is the
sound velocity of the phonon modes in units of the Fermi velocity of
the chiral states.  From Figure \ref{function.f} one sees that even
for moderately strong coupling between chiral edge state and phonons
the LL exponent $\nu$ is strongly renormalized. 
The system becomes unstable for coupling strengths $\gamma$ greater than a
critical value $\gamma_c$, which is equal to 0.56 for the 
example $u=0.61$ illustrated in the figure.

To calculate backscattered current and noise we use the nonequilibrium
Keldysh formalism. We define all fields on the Keldysh contour running
from $-\infty + i \epsilon$ to $\infty$ and then back to $-\infty - i
\epsilon$ \cite{Kleinert}. After Fourier transforming the Green function
Eq.(\ref{Greencross}) to imaginary time we analytically continue it to
the upper (lower) Keldysh contour by letting $\tau \to i t$ ($\tau \to
- i t$).  The Green functions for $q_\pm(t)\equiv 2 \sqrt{\pi}
\Theta(x=0,\tau\to \pm i t)$ at the impurity site are then given by

\begin{eqnarray}
  \label{eq:keldyshgreen}
 G_{\pm\pm}(t)&=&  \nu F \ln \left[ {\tau_c \pi T \over \sinh\left( 
t \pi T \right)}\right]^2 \, \mp \, 2 i \nu F \pi\\
G_{\pm\mp}(t)&=&\nu F\ln \left[ {\tau_c \pi T \over \sinh\left( t \pi T
\right)}\right]^2 \, \mp \, 2 i \nu F \pi {\rm sign}(t)\ \ ,\nonumber
\end{eqnarray}
with $\tau_c$ as a short time cutoff.  Now we can calculate the
partition function to second order in $\lambda$ and find

\begin{eqnarray}
  \label{eq:partitionfunction}
  \ln Z\left[\mu\right] &=&
{- \lambda^2 \over 2}\! \! \! \sum_{\alpha,\alpha^\prime=\pm}\!\!\!\! 
(\alpha \alpha^\prime)\! \! 
\int_{-\infty}^\infty\hspace*{-.5cm}  dt dt^\prime 
\left\langle\!\cos\!\left[\!q_\alpha(t)\! +\! \nu e V t \! +\! 
\nu e P_\alpha \mu(t)\!\right]
\right.\nonumber \\
\hspace{-1cm}& \hspace{-1cm} &\hspace*{-1cm} \left. \cos\left[q_{
\alpha^\prime}(t^\prime) + \nu e V t + \nu e 
P_{\alpha^\prime} \mu(t^\prime)\right]\right\rangle + \ln Z_0\left[\mu\right]  
\ .
\hspace*{-.5cm}
\end{eqnarray}
The expectation values are calculated with the Green functions
Eq.(\ref{eq:keldyshgreen}), $\mu(t)$ is a generating field for the
calculation of current correlation functions, and

\begin{eqnarray}
  \label{eq:projector}
  P_\alpha \mu (t) = \int {d\omega\over 2 \pi} e^{-i \omega t} \left({\alpha
\over 2} + \coth{\omega\over 2 T}\right) \mu(\omega)\ \ .
\end{eqnarray}

The backscattered current is given by $I_B=\nu e^2 V/h - \langle I \rangle$,
where the first term is the impinging current, and $\langle I \rangle$ is
the average transmitted current computed from $\langle I
\rangle = i \delta \ln Z / \delta \mu$. We thus find

\begin{eqnarray}
I_B  &=&R {e T\over \pi \hbar} \tan(\nu F \pi) {\Gamma(1\!-\!\nu F)\over \Gamma(\nu
F)} {\rm Im}\! \left[\!{\Gamma\left(\nu F \! -\! 
{i \nu e V  /2 \pi T}\right) \over \Gamma\left(1\! -\! \nu F \!-\! 
i \nu e V /2 \pi T\right)}\! \right] \nonumber\\
\label{phononcurrent}
\end{eqnarray}
with the reflection coefficient 
\begin{eqnarray}
R&=&\nu (\lambda \tau_c)^2 (\pi T \tau_c)^{2\nu F-2}{\pi^{3\over2}\over2}
\cos(\nu\pi F) \Gamma({1\over2}-\nu F)\Gamma(\nu F),\nonumber\\
\label{phononreflection}
\end{eqnarray}
which describes the backscattering probability in the linear transport regime.
As the microscopic parameters $\lambda$ and $\tau_c$ are not readily available
for experimental systems, it is useful to treat $R$ as an effective parameter
when comparing to experiments.
We define the noise as $\langle \Delta I(\omega) \Delta I(\omega^\prime)\rangle
= 2 \pi \delta(\omega + \omega^\prime) S(\omega)$. In the limit of 
$\omega$ going to zero we find

\begin{eqnarray}
  S&=& R T \nu {e^2 \over \pi \hbar}  {\Gamma(1\!-\!\nu F)\over \Gamma(\nu F)}
 {\rm Re}\left[{
\Gamma\left(\nu F\! - \!i\nu e V / 2\pi T\right)\over 
\Gamma\left(1\!-\!\nu F \!-\! i \nu e V  
/ 2\pi T\right)}\right]\nonumber\\
& &  -  2 T \left( 2 {\partial I_B\over \partial V} - \nu {e^2 \over 2 \pi
\hbar}
\label{phononnoise}
\right) \ .
\end{eqnarray}

In Figure 3, the total noise is plotted against the transmitted current for
a quasi-particle charge $e/3$ and different values of the
renormalization factor $F$. For $F=1$, we find LL behavior with
suppressed shot noise for large currents, as the impurity becomes more
and more transparent.  For a Fermi liquid (FL), which would correspond
to $F=3$, the shot noise grows linearly with the current. As an
example for an intermediate situation with a nonuniversal
noise--current characteristic we show the result for $F=2$.  

To test the accuracy of our perturbative calculation, we have compared
it for the special value $F=1$ (no renormalization of scattering
dynamics) to the exact solution in \cite{FeLuSa95,FeSa96}.  We find
good agreement for small reflection probabilities, e.g. for a
reflection coefficient $R=0.1$ the perturbative result for the shot
noise deviates less than 10\% from the exact solution.

\vspace*{.5cm}
\begin{figure}
\centerline{
\epsfig{file=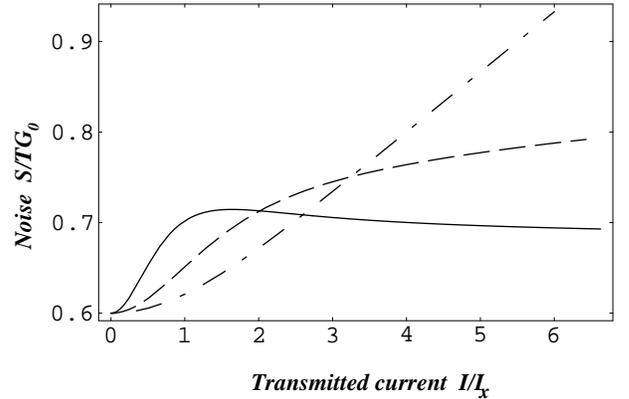,width=8cm}
}
\hspace*{.5cm}
\caption{Noise plotted against average transmitted current for weak 
  backscattering
  of charge $e/3$ quasi-particles with a linear reflection coefficient
  $R=0.1$. The noise is plotted in units of temperature times unit
  conductance $G_0 =e^2/h$, whereas the current is plotted in units of
  the crossover current $I_x=(1-R)e T/(\pi \hbar)$, which separates
  the regime of thermal Nyquist Johnson noise from the shot noise
  regime.  The full line corresponds to $\nu F=1/3$, the dashed line
  to $\nu F=2/3$ , and the dot dashed line to $\nu F=1$. This last case 
is the same as what we would obtain for a FL with particles of charge $e/3$.}
\label{comparison.f}
\end{figure}

We note that the coupling to the phonon mode changes the exponent
characterizing the scattering dynamics from $\nu$ to $\nu F$ in
Eqs.(\ref{phononcurrent},\ref{phononnoise}), while the fractional
charge $\nu e$ is not multiplied by $F$ and not renormalized.
As a consequence, in the
limit of dc voltage much larger than the temperature,
we find for the ratio of the shot noise to backscattered current

\begin{eqnarray}
  \label{eq:measurement}
{S- 2 T \nu {e^2 \over 2 \pi \hbar} \over I_B} =
\nu e   + O(T/V)\ .   
\end{eqnarray}

Thus the effective charge defined by this ratio coincides with the
fractional charge $\nu e$.

As stated above, we have ignored any effects of disorder or impurities at
the edge of the sample.  If there is disorder with Fourier
components commensurate with the reciprocal of the spacing between
electrons in the low-density edge strip, this can lead to pinning of the
Wigner crystal and an effective gap $\omega_g$ in the phonon dispersion 
relation.  (Equivalently, we may say that impurity-induced backscattering 
between the opposite moving edge modes of the low-density strip is relevant at 
low energies, and can lead to localization. Scattering of electrons
between the chiral edge state and the low-density strip should be
irrelevant at low energies, but it might also need to be taken into account
if the disorder is not sufficiently weak.)  If there is an effective energy
gap for the phonons, our analysis should still be valid,  provided that  $eV
\gg \hbar \omega_g$, while the shot noise is measured at frequencies low 
compared to $\omega_g$ and $\omega_L$.

How do the results for our model compare with experiments? In Ref.
\cite{PiReHeUm97}, at $\nu=1/3$, shot noise was found to correspond to
quasi-particles of charge e/3, while a linear I-V dependence was
observed over the voltage-range where the data was taken.  This result
could be understood directly in terms of our model if we assume an
appropriate value for the  coupling between a chiral edge-state and
additional phonon-like modes.  In Ref.  \cite{RePiGrHe99} , shot noise
corresponding to charge e/5 was observed (as expected) for the case of
$\nu=2/5$, while the I-V curves appear again to be more linear than
one would expect from a chiral LL model.  Since the case $\nu=2/5$
requires at least two chiral edge-modes, our model cannot be applied
directly; nevertheless, we expect that similar considerations apply.

The $\nu=1/3$ measurements of Ref. \cite{SaGlJi97}, by contrast, do
show a markedly nonlinear I-V characteristic, in the range where the
shot noise was measured.  Similarly, there appears to be a nonlinear
I-V characteristic in the data exhibited in Ref.\cite{GrCoHe00} at
$\nu=2/5$. 
In Ref. \cite{Gl00} the nonlinearity was compared explicitly with
the LL predictions and found to deviate in the regime of small
temperatures and low voltages where backscattering is strong,
but there is no quantitative comparison in the regime where the 
backscattering is weak. 
In all these  cases the data were taken over an extended range of the
reflectivity R, not just in the limit of weak back scattering, and the
shot noise was analyzed using a formula with an additional factor
(1-R), typical of non-interacting fermions. As our calculation is only
valid to lowest-order in R, we cannot make a statement about the
appearance of such a factor.

In summary, we have studied the dynamics of scattering between
fractional quantum Hall edges in the presence of additional phonon
like edge modes. We find that the coupling between these additional
modes and the current carrying edge mode leads to a nonuniversal
current-voltage characteristic which may explain the deviations from 
Luttinger liquid behavior observed in experiments. The fractional charge
of FQHE quasi-particles, however, can still be measured as the quotient
of shot noise and backscattered current, in the limit of weak backscattering. 

{\it Acknowledgements:} We acknowledge helpful conversations with 
A. Stern, D.C. Glattli, M. Reznikov, M. Heiblum, Y. Oreg, and S. Kehrein.  
This work was supported in part by NSF grant
DMR99-81283. B.R. thanks DFG for financial support under grant
Ro2247/1-1.



\end{multicols}

\end{document}